\documentclass[a4paper,fleqn,usenatbib]{mnras}

\usepackage[T1]{fontenc}
\usepackage{ae,aecompl}

\usepackage{graphicx}	
\usepackage{amsmath}	
\usepackage{amssymb}	

\title[Rayleigh scattering in HAT-P-18b]{Rayleigh scattering in the transmission spectrum of HAT-P-18b}

\author[J. Kirk et al.]{
J. Kirk,$^{1}$\thanks{E-mail: James.Kirk@warwick.ac.uk}
P. J. Wheatley\thanks{E-mail: P.J.Wheatley@warwick.ac.uk},$^{1}$
T. Louden,$^{1}$
A. P. Doyle,$^{1}$
I. Skillen,$^{2}$ 
J. McCormac,$^{1}$ \newauthor
P. G. J. Irwin,$^{3}$
and R. Karjalainen$^{2}$
\\
$^1$Department of Physics, University of Warwick, Coventry, CV4 7AL, UK \\
$^{2}$Isaac Newton Group of Telescopes, Apartado de Correos 321, 38700 Santa Cruz de Palma, Spain\\
$^{3}$Atmospheric, Oceanic, and Planetary Physics, Clarendon Laboratory, University of Oxford, Parks Road, Oxford OX1 3PU, UK\\
}

\date{Accepted XXX. Received YYY; in original form ZZZ}

\pubyear{2016}

\begin{document}
\label{firstpage}
\pagerange{\pageref{firstpage}--\pageref{lastpage}}
\maketitle

\begin{abstract}
We have performed ground-based transmission spectroscopy of the hot Jupiter HAT-P-18b using the ACAM instrument on the William Herschel Telescope (WHT). Differential spectroscopy over an entire night was carried out at a resolution of $R \approx 400$ using a nearby comparison star. We detect a bluewards slope extending across our optical transmission spectrum which runs from 4750\,\AA ~to 9250\,\AA. The slope is consistent with Rayleigh scattering at the equilibrium temperature of the planet (852\,K). We do not detect enhanced sodium absorption, which indicates that a high-altitude haze is masking the feature and giving rise to the Rayleigh slope. This is only the second discovery of a Rayleigh scattering slope in a hot Jupiter atmosphere from the ground, and our study illustrates how ground-based observations can provide transmission spectra with precision comparable to the \emph{Hubble Space Telescope}.
\end{abstract}




\section{Introduction}

Transmission spectroscopy is an essential tool to characterise the atmospheres of transiting exoplanets (e.g. \citealt{Charbonneau2002}; \citealt{Redfield2008}; \citealt{Snellen2008}) and the technique is particularly well suited to hot Jupiters due to their large atmospheric scale heights. The sample of hot Jupiters studied to date is revealing a diverse array of atmospheric chemistries which often depart from models of clear atmospheres, which show strong alkali metal absorption and H$_2$ induced Rayleigh scattering slopes towards the blue optical (\citealt{SeagerSasselov}; \citealt{Brown2001}). To date only a single hot Jupiter has been found to display the broad wings of the sodium and potassium features (WASP-39b; \citealt{Fischer2016}; \citealt{Sing2016}). Increasingly, transmission spectroscopy studies reveal cloud- and haze-dominated atmospheres that mask the broad wings of sodium and potassium (e.g. \citealt{Pont2013}; \citealt{Nikolov2014}), whilst in some cases even the narrow cores are not seen (e.g. \citealt{Sing2013}). The range of transmission spectra hot Jupiters display was highlighted in a recent survey of 10 hot Jupiters with the \emph{Hubble Space Telescope} (\emph{HST}) and \emph{Spitzer Space Telescope} \citep{Sing2016}. This revealed a continuum in hot Jupiter atmospheres from clear atmospheres showing strong alkali metal absorption to cloud- and haze-dominated atmospheres muting, and sometimes masking entirely, the alkali absorption features.

Ground-based detections of spectral features within hot Jupiter atmospheres have predominantly been of the narrow line cores of sodium (e.g. \citealt{Snellen2008}; \citealt{Zhou2012}; \citealt{Nikolov2016}) and potassium (e.g. \citealt{Wilson2015}; \citealt{Sedaghati2016}). Whilst there have been a couple of detections of bluewards scattering slopes from the ground (e.g. \citealt{Jordan2013}; \citealt{DiGloria2015}), ground-based measurements of hot Jupiters have often revealed featureless transmission spectra dominated by clouds (e.g. \citealt{Kirk2016}; \citealt{Mallonn2016}). The diversity of these results has emphasised the need to increase the current pool of studied gas giant atmospheres to better understand the processes and parameters governing the presence or absence of clouds and hazes.

In this paper, we present the low-resolution ground-based transmission spectrum of the Saturn-mass planet HAT-P-18b \citep{Hartman2011}. HAT-P-18b's equilibrium temperature of 852\,K is relatively cool for a hot Jupiter owing to its relatively long period of 5.5\,d. HAT-P-18b's inflated radius (0.995\,$R_\text{J}$) and relatively low mass (0.197\,$M_\text{J}$) lead to a large atmospheric scale height of 540\,km. The combination of its large scale height with the small radius of the host star (0.749\,R$_{\odot}$) makes HAT-P-18b well suited to transmission spectroscopy.

\section{Observations}

HAT-P-18 was observed on the night of 2016 April 28 with the ACAM instrument \citep{Benn2008} on the 4.2-m William Herschel Telescope (WHT), La Palma. ACAM\footnote{http://www.ing.iac.es/astronomy/instruments/acam/}  is mounted at the folded-Cassegrain focus of the WHT and can be used both for broad- and narrow-band imaging and for low-resolution spectroscopy using a Volume Phase Holographic disperser. For this work, ACAM was used in spectroscopy mode, providing low resolution spectroscopy over the entire optical range with a throughput ranging between 0.5 and 0.8 \citep{Benn2008}.

ACAM was used unbinned in fast readout mode with a smaller than standard detector window to reduce the overhead to $\sim11$\,s with exposure times of 60\,s. This integration time was used to limit the peak counts of the comparison star to the range of the CCD response characterised as linear\footnote{http://www.ing.iac.es/Engineering/detectors/auxcam\_fast.jpg}. A custom made 27 arcsec slit was used to perform these observations to avoid the harmful effects of differential slit losses between the target and its comparison star which can lead to systematics in the derived transmission spectra. To further avoid slit losses, the locations of the spectral traces were monitored throughout the night, and manual guiding corrections made to keep the traces within a couple of pixels in the x (spatial) and y (dispersion) directions respectively (Fig. \ref{fig:combined_plots}). A slight defocus was used to maintain the full width at half-maximum at around 1 arcsec. A total of 320 spectra were observed over the course of the night with airmass varying from 1.93 to 1.00 to 1.08 and a moon illumination of 62 per cent. Biases, sky flats and spectra of the CuAr plus CuNe arcs were taken at the start and end of the night.

A comparison star with a similar magnitude to HAT-P-18 was used to perform differential spectroscopy in order to account for telluric variability. The comparison star used was TYC 2594-731-1 at a distance of 3.4 arcmin from HAT-P-18 with a $V$ magnitude of 11.2 and $B-V$ colour of 1.3. HAT-P-18 has a $V$ magnitude of 12.7 and $B-V$ colour of 1.0. The separation between the stars was comfortably within ACAM's slit length of 6.8 arcmin. The comparison star is not known to be a variable star.

\section{Data Reduction}

To reduce the data, \textsc{python} scripts were written from scratch to create biases, flat-fields and to extract the spectral traces. 51 bias frames were median-combined to create a master bias which was subsequently subtracted from each of the science frames. We tried several methods to flat field but found the least red noise in the white light curve resulted from no flat-fielding. Lamp flats did not contain enough blue photons while sky flats naturally contained structure. We tried slitless sky flats but these were contaminated by a sector of the zero-order image. Since the spectral traces were spread across a large number of pixels, the lack of flat-fielding did not form a significant source of error.

\begin{figure}
\centering
\includegraphics[scale=3.5]{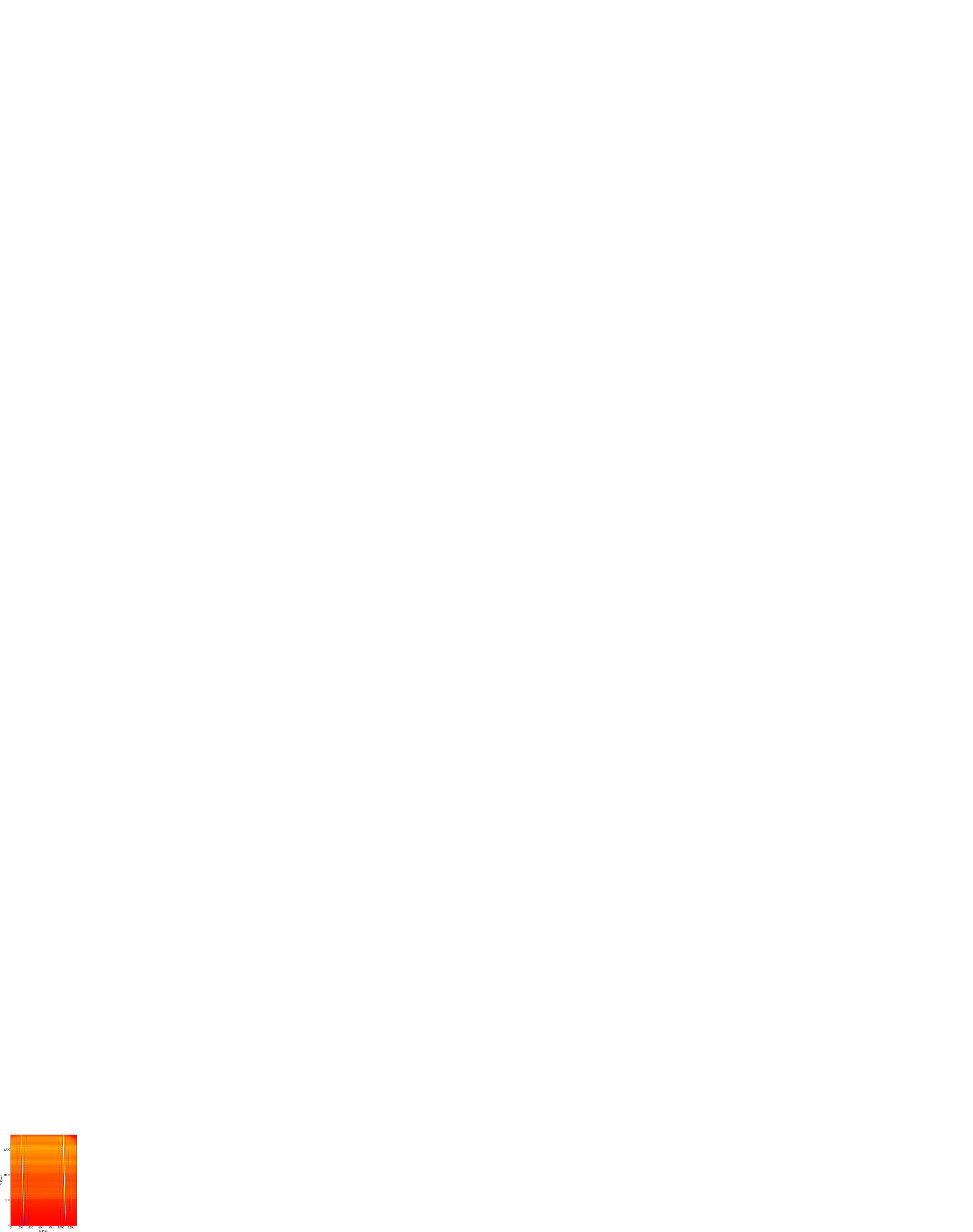}
\caption{Example frame with extraction regions for the target (left-hand trace) and comparison (right-hand trace). The solid lines indicate the extraction apertures with the dotted lines indicating the sky background regions.}
\label{fig:frame}
\end{figure}

To extract the spectra, polynomials were fitted along each of the two traces in the dispersion direction and apertures placed around these. Fig. \ref{fig:frame} shows an example frame with the aperture and background regions overplotted and Fig. \ref{fig:bin_locations} shows an example extracted spectrum. An iterative polynomial background fit was performed row by row across the combined width of the extraction and background apertures and the resulting polynomial subtracted. The iterative nature of this means that the spectral traces themselves are naturally masked from the resulting polynomial, which is, instead, just modelling the background variation in the spatial direction. After subtraction of the polynomial, normal extraction of the counts within the aperture was performed. A number of different extraction and background aperture widths were experimented with to minimise the scatter in the resulting light curves. This resulted in an optimal aperture width of 30 pixels. The background was calculated from two 50 pixel-wide regions, either side of the aperture, at distances of 30 pixels from the edges of the aperture. The pixel scale of ACAM is 0.25 arcsec pixel$^{-1}$. The errors in the data points were a combination of the photon noise and the readnoise of ACAM, which, in fast readout mode, is 7 electrons pixel$^{-1}$ with a gain of 1.9 electrons count$^{-1}$.

After extraction of the spectral traces, cosmic rays were removed. This was done by first dividing each spectral frame by a reference frame clean of cosmics. This self-division of each spectral trace resulted in spikes where cosmic rays were located. The flux values of the pixels affected by cosmic rays were then replaced by interpolating between the two nearest neighbouring pixels unaffected by the cosmic ray.

The differential white light curve of the target is plotted in Fig. \ref{fig:combined_plots} along with airmass and diagnostics of the traces over the course of the observation. This figure shows the small rotation of the traces on the CCD resulting from a combination of instrument flexure \citep{Benn2008} and atmospheric refraction as there was no atmospheric dispersion corrector. Due to this rotation, the spectra were aligned in wavelength so that accurate differential spectroscopy could be performed. This was done by cross-correlating regions of each spectral frame with a reference frame around significant absorption features. This was repeated across the spectrum and allowed for the shift relative to the reference to be calculated as a function of location on the chip. A third order polynomial was then fitted to these shifts and the individual spectra resampled onto the grid of the reference spectrum. The shifts for each star's spectra were calculated individually before resampling onto the same grid. This resulted in the spectra of both stars being well aligned for all frames.

With the spectra aligned in pixel space, the wavelength solution was calculated using strong telluric and stellar absorption lines. A synthetic spectrum of telluric and spectral lines with a resolution of $R = 1000$ was generated for a star of the same temperature, surface gravity and metallicity as HAT-P-18 which was cross-correlated with the observed spectrum. This gave the wavelength as a function of pixel position for a number of strong features. A second-order polynomial was then fitted to convert from pixel position into wavelength. We also calculated the arc solution using arcs taken through a 1 arcsec slit and found that these two solutions were consistent to within a couple of angstroms (1 per cent of the bin width).

For further analysis the very blue and very red ends of the spectra were excluded due to low signal to noise and differential vignetting between the target and comparison. This resulted in the spectra spanning a wavelength range between 4750\,\AA ~and 9250\,\AA. 

\section{Data Analysis}

\subsection{Light curve fitting with free limb darkening}
\label{sec:free_lds}

\begin{figure}
\centering
\includegraphics[scale=0.25]{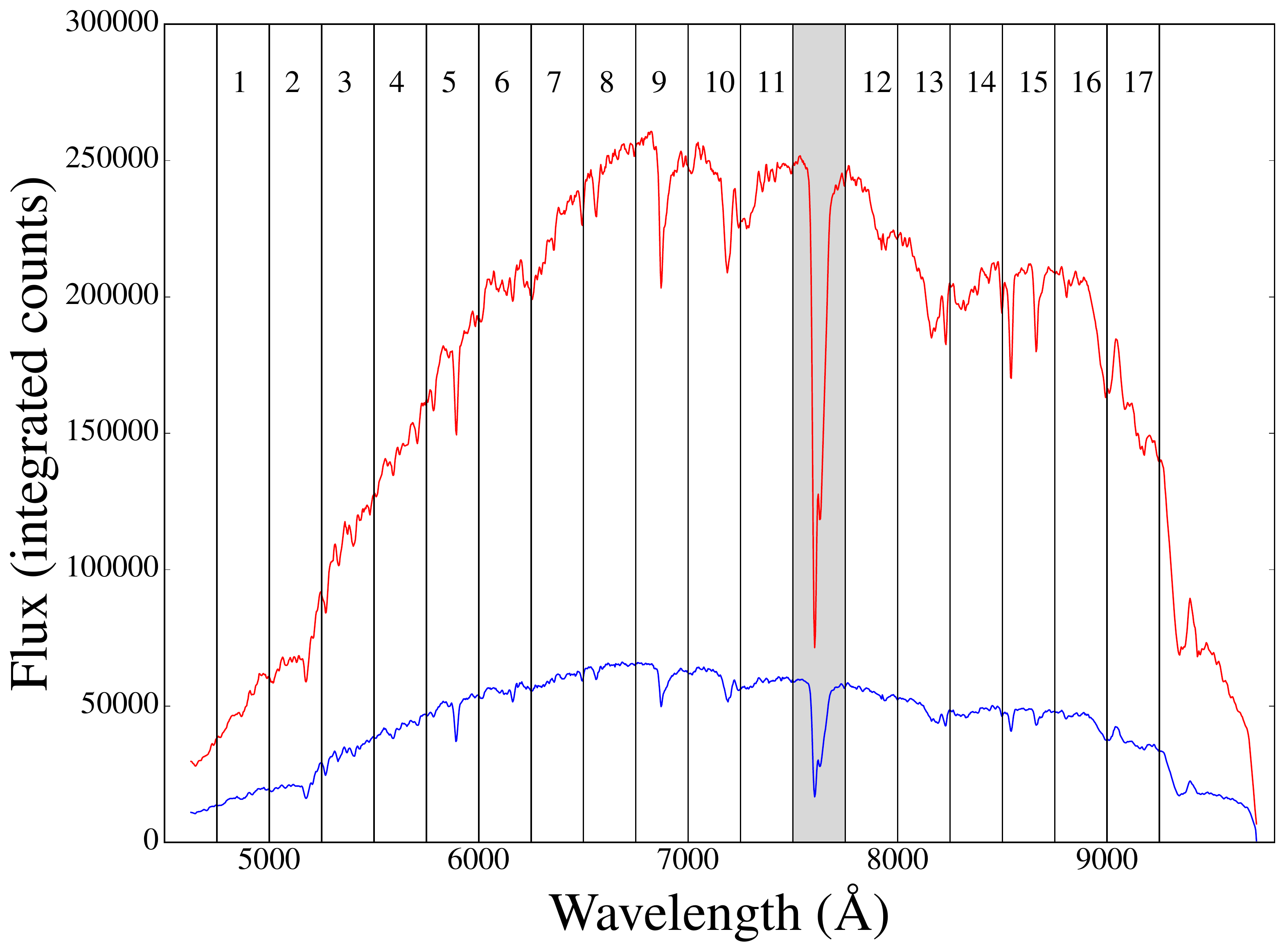}
\caption{Locations of the seventeen 250\,\AA -wide bins used in the wavelength bin fitting. The target's spectrum is shown in blue and the comparison's in red, in raw counts for a single exposure. The grey region shows the bin containing the strong telluric feature that was excluded from further analysis.}
\label{fig:bin_locations}
\end{figure}

With the data reduced and spectra extracted, we binned the data into 250\,\AA -wide wavelength bins running from 4750\,\AA ~to 9250\,\AA. This bin width was chosen as a compromise between resolution and signal-to-noise ratio. The bin containing the strong telluric oxygen feature at $\sim7600$\,\AA ~showed a significant level of red noise and so this bin was masked from further analysis. This resulted in seventeen 250\,\AA -wide bins\footnote{The reduced light curves are available with the online version of this paper.}. The bin locations are shown in Fig. \ref{fig:bin_locations} with the masked region shaded grey.

We fitted the light curves with analytic transit light curves using a quadratic limb darkening law \citep{MandelAgol}. A long time-scale trend was simultaneously fitted to model the small overall trend which was less than a mmag in amplitude. This trend was most likely a second-order colour extinction effect in the 250\,\AA -wide bins. We experimented with different functions to fit this trend, including: a quadratic in time polynomial, a cubic in time polynomial, a linear in time extinction coefficient and a quadratic in time extinction coefficient. All functions resulted in transmission spectra with blueward slopes, and the cubic in time polynomial was adopted for our final results due to the better Bayesian information criterion (BIC; \citealt{Schwarz1978}). The parameters defining the model were the radius ratio of planet to star $R_P/R_*$, the scaled stellar radius $a/R_*$, the inclination $i$, the quadratic limb darkening coefficients $u1$ and $u2$, the time of mid-transit $T_c$, and the four parameters defining the long time-scale trend. 

The light curve models were initially fitted using \textsc{scipy}'s \textsc{optimize} package within \textsc{python} \citep{scipy} using a Nelder-Mead algorithm to perform the minimisation \citep{NelderMead}. With the results from this fit a Markov chain Monte Carlo (MCMC) was initiated at these values and was implemented using the \textsc{python} package \textsc{emcee} \citep{emcee}.

\begin{figure}
\centering
\includegraphics[scale=0.25]{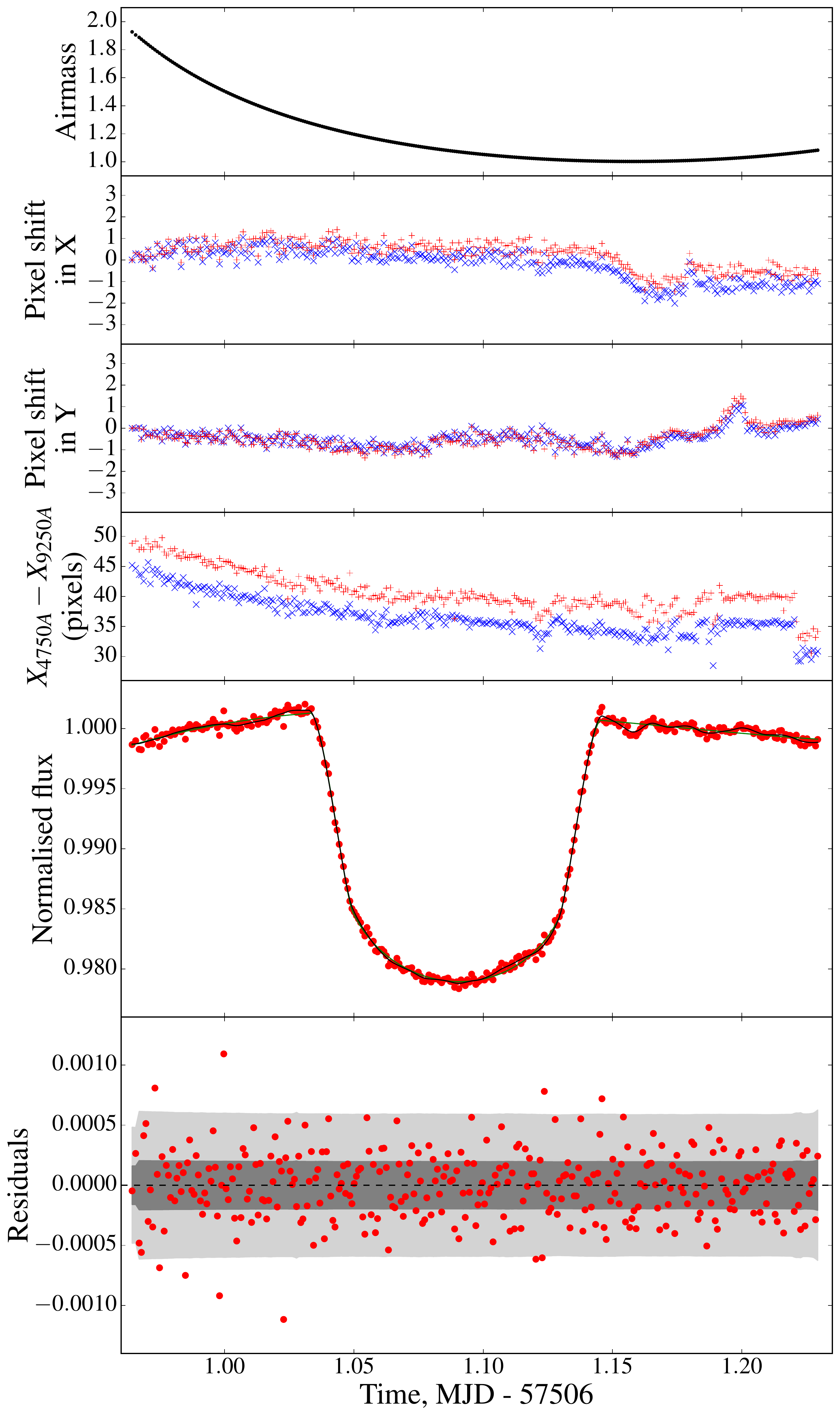}
\caption{Top panel: Variation of airmass across the night. Second panel: Shift in the pixel positions in x (spatial direction) for the target trace (blue crosses) and comparison trace (red pluses). Third panel: Shift in the pixel positions in y (dispersion direction) for the target trace (blue crosses) and comparison trace (red pluses). Fourth panel: The rotation of the target trace (blue crosses) and comparison trace (red pluses) shown as the difference in x position at the bottom and top of the trace. Fifth panel: The raw white light curve shown by red data points and generated from summing up the bins shown in Fig. \protect\ref{fig:bin_locations}. Over plotted are the fits from the analytic transit light curve with a cubic in time polynomial (green line) and a Gaussian process (GP, black line). Sixth panel: Residuals to the GP fit are given by the red points with the dark grey and light grey shaded regions indicating the 1$\sigma$ and 3$\sigma$ confidence intervals, respectively.}
\label{fig:combined_plots}
\end{figure}

The system parameters ($a/R_*$, $i$ and $T_c$) were held fixed to the results from a fit to the white light curve (Fig. \ref{fig:combined_plots}, Table \ref{tab:system_params}). This was created simply by summing the seventeen individual wavelength-binned light curves. We did this as we are interested in the relative error in the planetary radius between wavelengths and not in the absolute error in the radius. By fixing these system parameters shared between the wavelength bins, we remove them as sources of error within the relative radii.

We initially fitted the light curves with the quadratic limb darkening coefficients $u1$ and $u2$ as free parameters. We used a uniform prior such that $u1 +u2 \leq 1$ and placed no priors on the other parameters. With the starting values resulting from the Nelder-Mead minimisation, an initial MCMC to each wavelength-binned light curve was run for 1000 iterations with 140 walkers ($20 \times n_p$, where $n_p$ is the number of parameters). The first 500 steps were discarded as the chains were burning in. The best-fitting model resulting from this initial run was then used to rescale the error bars to give $\chi^2_{\nu} = 1$ for each wavelength bin. Following this, a second MCMC was run with 140 walkers and for 1000 iterations.

\begin{figure*}
\includegraphics[scale=0.5]{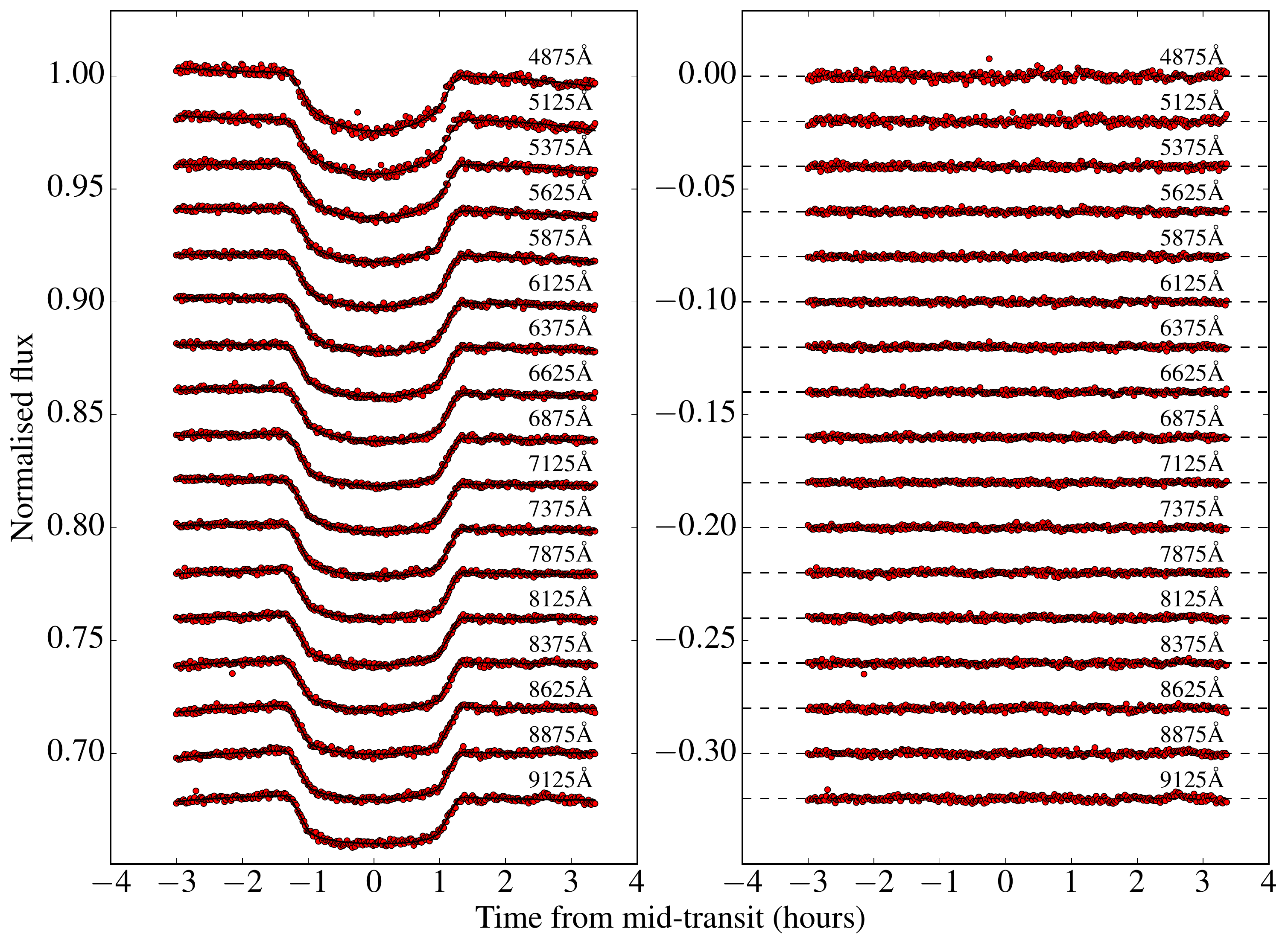}
\caption{Left-hand panel: Our fits to the wavelength binned light curves of HAT-P-18b, going from blue (top) to red (bottom), which are offset for clarity. The black lines are the fitted models with both limb darkening coefficients as fit parameters and using the cubic in time polynomial, giving 7 free parameters per wavelength bin. The red points show the normalised differential flux without any detrending. The quoted wavelengths are those of the centre of each bin. Right-hand panel: the residuals to the fits in the left-hand panel after subtracting the best-fitting model.}
\label{fig:bin_fits}
\end{figure*}

\begin{figure*}
\centering
\includegraphics[scale=0.5]{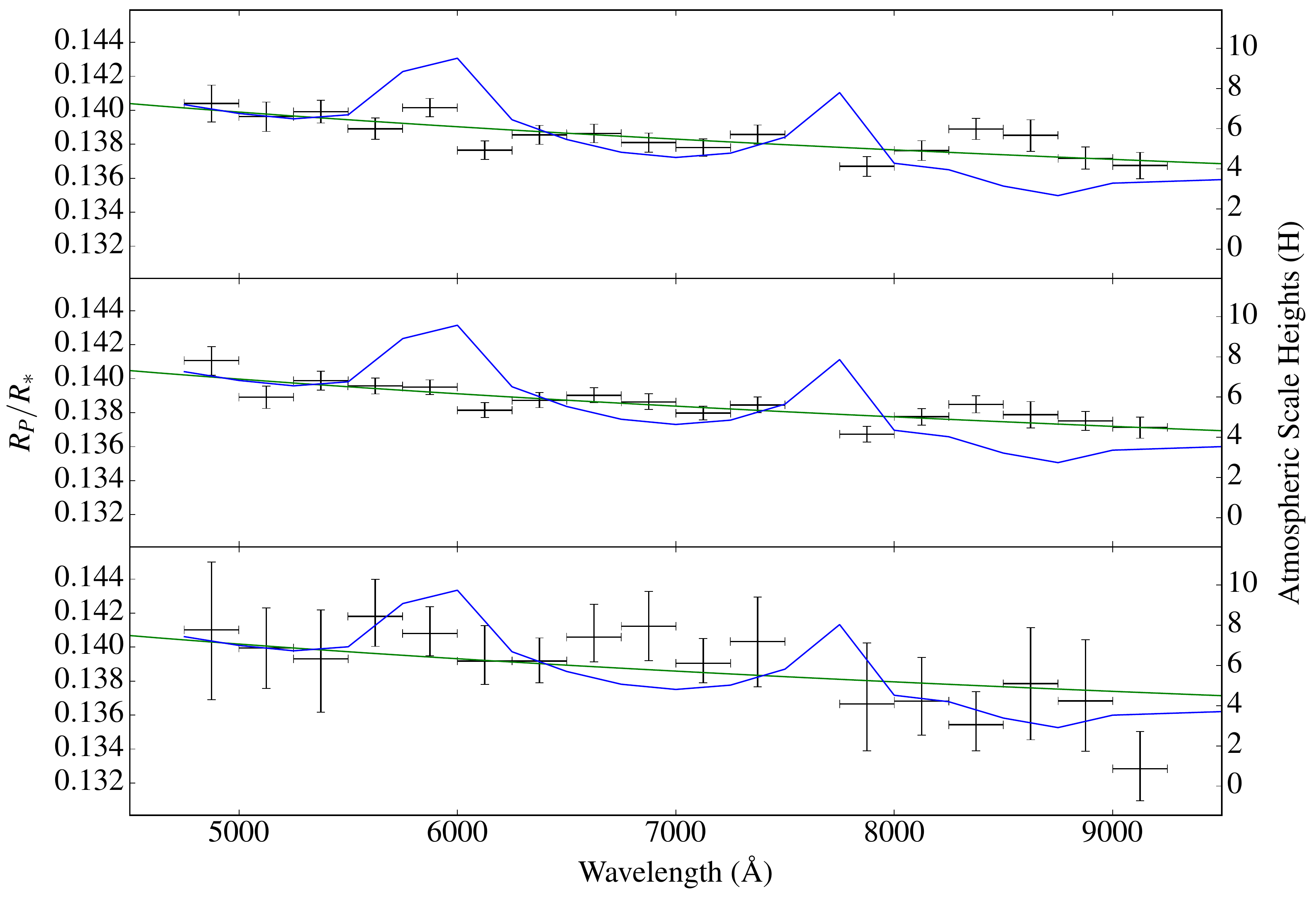}
\caption{Transmission spectrum of HAT-P-18b resulting from different treatments of the limb darkening and long time scale trend. Top panel: analytic transit light curves with a cubic in time polynomial. In this fit, both $u1$ and $u2$ were free parameters. Second panel: analytic transit light curves with a cubic in time polynomial. In this fit $u1$ was fixed with $u2$ described by a polynomial. Bottom panel: analytic transit light curves with a Gaussian process. In this fit $u1$ was fixed with $u2$ described by a polynomial. In each panel, the black data points are the resulting $R_P/R_*$ values. The green line shows a Rayleigh scattering slope at the equilibrium temperature of the planet (852\,K). The blue line shows a clear atmosphere model, binned to the size of the data points, highlighting the absence of the broad wings of the Na and K features, indicating the presence of a high-altitude haze.}
\label{fig:trans_spec}
\end{figure*}

The results of the fits to the wavelength-binned light curves are shown in Fig. \ref{fig:bin_fits} and Table \ref{tab:bin_fits} with the rms of the residuals between 20 per cent and 80 per cent of the expected photon precision. The resulting transmission spectrum displayed a gradient running from blue to red (Fig. \ref{fig:trans_spec}, top panel).

\begin{table*}
\caption{Results from the fitting of the wavelength binned light curves shown in Fig. \protect\ref{fig:bin_fits}.}
\label{tab:bin_fits}
\centering
\begin{tabular}{l c c c c}\hline
Bin & $R_P/R_*$ & $u1$ & $u2$ & Residual rms (mmag) \\ \hline

4750 -- 4999\,\AA & $0.14048^{+0.00108}_{-0.00110}$ & $0.92^{+0.11}_{-0.12}$ & $-0.142^{+0.186}_{-0.177}$ & 1.44 \\
5000 -- 5249\,\AA & $0.13970^{+0.00085}_{-0.00087}$ & $1.00 \pm 0.10$ & $-0.382^{+0.155}_{-0.154}$ & 1.18 \\
5250 -- 5499\,\AA & $0.13999 \pm 0.00067$ & $0.81 \pm 0.08$ & $-0.132^{+0.121}_{-0.123}$ & 0.91 \\
5500 -- 5749\,\AA & $0.13898^{+0.00064}_{-0.00061}$& $0.55 \pm 0.07$ & $0.198^{+0.111}_{-0.107}$ & 0.78 \\
5750 -- 5999\,\AA & $0.14023^{+0.00054}_{-0.00053}$& $0.63 \pm 0.07$ & $-0.035^{+0.100}_{-0.103}$ & 0.70 \\
6000 -- 6249\,\AA & $0.13771^{+0.00056}_{-0.00055}$& $0.48\pm 0.07$ & $0.203^{+0.109}_{-0.104}$ & 0.69 \\
6250 -- 6499\,\AA & $0.13862 \pm 0.00056$ & $0.61\pm 0.07$ & $-0.030^{+0.107}_{-0.104}$ & 0.72 \\
6500 -- 6749\,\AA & $0.13870^{+0.00054}_{-0.00055}$& $0.50\pm 0.07$ & $0.104\pm0.104$ & 0.69 \\
6750 -- 6999\,\AA & $0.13817^{+0.00056}_{-0.00057}$& $0.41^{+0.07}_{-0.08}$ & $0.206^{+0.112}_{-0.110}$ & 0.70 \\
7000 -- 7249\,\AA & $0.13787 \pm 0.00051$ & $0.48\pm 0.07$ & $0.067^{+0.099}_{-0.102}$ & 0.64 \\
7250 -- 7499\,\AA & $0.13863 \pm 0.00056$ & $0.48^{+0.07}_{-0.08}$ & $0.018^{+0.112}_{-0.109}$ & 0.70 \\
7750 -- 7999\,\AA & $0.13676^{+0.00058}_{-0.00059}$& $0.43\pm 0.08$ & $0.077^{+0.118}_{-0.116}$ & 0.73 \\
8000 -- 8249\,\AA & $0.13769^{+0.00058}_{-0.00059}$& $0.60^{+0.07}_{-0.08}$ & $-0.138^{+0.113}_{-0.111}$ & 0.75 \\
8250 -- 8499\,\AA & $0.13895 \pm 0.00062$ & $0.58\pm 0.08$ & $-0.159^{+0.115}_{-0.112}$ & 0.77 \\
8500 -- 8749\,\AA & $0.13858^{+0.00091}_{-0.00095}$& $0.47^{+0.12}_{-0.13}$ & $-0.031^{+0.187}_{-0.185}$ & 1.20 \\
8750 -- 8999\,\AA & $0.13722^{+0.00067}_{-0.00063}$& $0.44\pm 0.09$ & $0.090^{+0.133}_{-0.130}$ & 0.84 \\
9000 -- 9249\,\AA & $0.13680^{+0.00077}_{-0.00078}$& $0.27^{+0.10}_{-0.11}$ & $0.328^{+0.165}_{-0.153}$ & 0.96 \\ \hline

\end{tabular}
\end{table*}

\subsection{Light curve fitting with constrained limb darkening}
\label{sec:constrained_lds}

\begin{figure}
\centering
\includegraphics[scale=0.25]{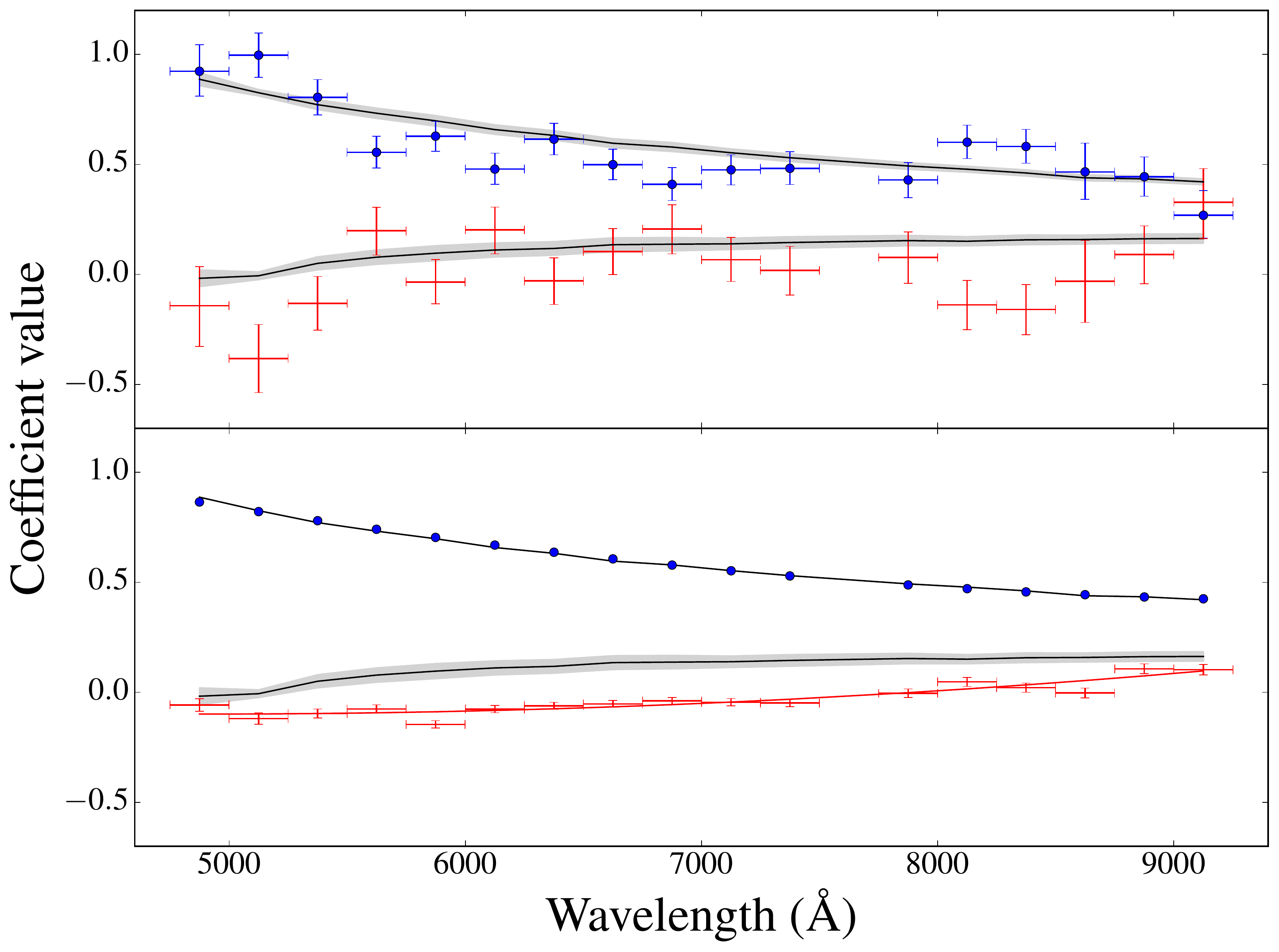}
\caption{Top panel: plot of the limb darkening coefficients when both are fitted as free parameters. The blue error bars show the fitted values for $u1$ and the red error bars show those for $u2$. The solid black lines show the predicted values for $u1$ and $u2$ generated by \textsc{ldtk} while the grey regions show the 1$\sigma$ confidence regions. Bottom panel: results when $u1$ is held fixed to a polynomial fitted to the predicted values of $u1$. The blue data points indicate the values that $u1$ is fixed to for each bin. The red error bars indicate the fitted values for $u2$ with the solid red line showing the polynomial fitted to these values of $u2$. The black line and grey regions are the same as in the top panel.}
\label{fig:ldcs}
\end{figure}

When fitting for both limb darkening coefficients, we found the resulting values were consistent with predicted values for this star (Fig. \ref{fig:ldcs}). To generate the predicted values, we made use of the limb darkening toolkit (\textsc{ldtk}, \citealt{LDTK}) which uses \textsc{phoenix} models \citep{Husser2013} to calculate $u1$ and $u2$ with errors propagated from the errors in the stellar parameters. Fig. \ref{fig:ldcs} also shows the anticorrelation between $u1$ and $u2$. Because of this degeneracy, it is common to hold one coefficient fixed and fit for the other (e.g. \citealt{Southworth2008}; \citealt{Kirk2016}) and so we experimented with this also. Since we are interested in the relative radii between wavelengths, fixing one of the limb darkening coefficients removes this as a source of error within $R_P/R_*$. We therefore ran another MCMC with $u1$ fixed and $u2$ as a free parameter.

We held $u1$ fixed to a second-order polynomial fitted to the predicted values of $u1$ generated by \textsc{ldtk} (Fig. \ref{fig:ldcs}, lower panel). This removed $u1$ as a fit parameter, leaving $R_P/R_*$, $u2$, and the four parameters defining the long time-scale trend as fit parameters. After an initial MCMC to each wavelength binned light curve was run for 1000 iterations with 120 walkers ($20 \times n_p$), with the first 500 steps discarded as the chains were burning-in, the best fitting model was used to rescale the error bars giving $\chi^2_{\nu} = 1$ for each wavelength bin. Following this initial run of the MCMC, we fitted a second-order polynomial to the resulting values for $u2$ to describe $u2$ as a function of wavelength (Fig. \ref{fig:ldcs}, lower panel). A second MCMC was then run with 100 walkers ($20 \times n_p$) for 1000 steps with this function describing $u2$, thus removing it as fit parameter.

This treatment of the limb darkening led to the same slope in the transmission spectrum as the fit with the limb darkening coefficients free, but with slightly smaller errors in $R_P/R_*$ (Fig. \ref{fig:trans_spec}, middle panel). Fig. \ref{fig:ldcs} (lower panel) shows the fitted values of $u2$ compared with those generated by \textsc{ldtk}. While the fitted values for $u2$ did differ to those from \textsc{ldtk}, importantly, this did not affect the slope in the transmission spectrum. The difference between the predicted and fitted limb darkening coefficients resulting from this method is discussed in section \ref{sec:limb_darkening}.

\subsection{Light curve fitting with Gaussian process detrending}

Whilst the models using the cubic in time polynomial provided good fits to the data (Fig. \ref{fig:bin_fits}), we wanted to confirm that the transmission spectra were independent of the function used to model the long time-scale trend. To do this, we used a Gaussian process (GP) which is a non-parametrised method to model covariance in data. This offered a robust test of whether the results from the use of the cubic in time polynomial were independent of the choice of function used. GPs have been demonstrated to be powerful in modelling correlated noise in exoplanet transit light curves by, for example, \cite{Gibson2012,Gibson2013_hat32} and \cite{Evans2015}.

The GP was implemented through the \textsc{george} \textsc{python} package \citep{george}. The mean function of the GP in each wavelength bin was the quadratically limb darkened analytic transit light curve \citep{MandelAgol}. We treated the limb darkening in the same way as described in section \ref{sec:constrained_lds}. We used a Mat\'{e}rn 3/2 kernel to model correlations in the data, defined by the hyper-parameters $\tau$ (the time-scale) and $a$ (the amplitude), in addition to a white-noise kernel defined by the variance $\sigma$. The fit parameters were therefore $R_P/R_*$, $u2$, $a$, $\tau$ and $\sigma$, with $u1$ described by the same polynomial as before.  We used loose, uniform priors to encourage convergence.

An initial MCMC was run for 2000 steps with 100 walkers ($20 \times n_p$) with the first 1000 steps discarded as burn-in. As before, we then ran a second MCMC with a function describing the fitted values of $u2$ resulting from the first run, thus removing $u2$ as a fit parameter. The second run was 2000 steps long with 80 walkers ($20 \times n_p$).

When fitting with the GP, we held the system parameters ($i$, $a/R_*$ and $T_C$) fixed to values obtained from fitting the white light curve with a GP (Fig. \ref{fig:combined_plots}, Table \ref{tab:system_params}).

\subsection{Transmission spectrum}

The transmission spectrum resulting from each of the three fitting methods is displayed in Fig. \ref{fig:trans_spec}, revealing a strong blueward slope resembling a Rayleigh scattering signature. The transmission spectra were consistent between all methods, confirming that the presence of the slope was independent of the choice of function used to fit the long time scale trend and whether we fixed or fitted the limb darkening coefficients. The GP resulted in larger errors in the transmission spectrum due to its ability to model a much wider set of detrending functions.

A Rayleigh scattering slope is plotted on this figure at the equilibrium temperature of the planet (852\,K; \citealt{Hartman2011}), with a slope given by \cite{Etangs2008_HD189} as

\begin{equation}
\frac{dR_{p}}{d\ln\lambda} = \frac{k}{\mu g} \alpha T
\end{equation}
\noindent
where $\mu$ is the mean molecular mass of an atmospheric particle taken to be 2.3 times the mass of a proton, $k$ is the Boltzmann constant, $g$ is the planet's surface gravity, $\alpha = -4$ as expected for Rayleigh scattering, and $T$ we take as the equilibrium temperature.

The transmission spectra resulting from the three different modelling techniques were all well fitted by a Rayleigh slope at the equilibrium temperature of the planet (852\,K).

When using the cubic in time polynomial with both $u1$ and $u2$ as free parameters (Fig. \ref{fig:trans_spec}, top panel) the $\Delta$BIC between a Rayleigh slope at the equilibrium temperature and a flat transmission spectrum was 22.9. This very strongly favoured the Rayleigh slope \citep{KassRaferty}. A fit to this transmission spectrum with the temperature as a free parameter resulted in a temperature of $784 \pm 194$\,K, consistent with the equilibrium temperature.

When fixing $u1$ and using the cubic in time polynomial (Fig. \ref{fig:trans_spec}, middle panel), the $\Delta$BIC between the Rayleigh slope at the equilibrium temperature and a flat transmission spectrum was 37.1, which again very strongly favoured the Rayleigh slope. The temperature resulting from a Rayleigh slope fitted to the transmission spectrum was $798 \pm 150$\,K, which was again consistent with the equilibrium temperature.

The GP resulted in larger errors in the transmission spectrum (Fig. \ref{fig:trans_spec}, bottom panel) and consequently lower values of the $\Delta$BIC between models. However, a Rayleigh slope at the equilibrium temperature was still strongly favoured over a flat transmission spectrum with a $\Delta$BIC of 8.6. When fitting this transmission spectrum with a free temperature, the preferred gradient was at a significantly higher temperature of $2023 \pm 393$\,K. However, the $\Delta$BIC between the fitted temperature and equilibrium temperature was less than 2, which is insignificant, and so we do not conclude this as evidence for a temperature inversion.

We also plot on Fig. \ref{fig:trans_spec} a clear atmosphere model (in blue) as resulting from the \textsc{nemesis} radiative transfer code \citep{irwin2008nemesis}, binned to the resolution of the data. The clear atmosphere model does not provide as good a fit as the Rayleigh scattering slope, as we do not detect the broad wings of the strong sodium and potassium features. This indicates a condensate haze is masking the wings as in, for example, HD~189733b (\citealt{Pont2008}; \citealt{Sing2011}; \citealt{Huitson2012}; \citealt{Pont2013}), WASP-31b \citep{Sing2015} and WASP-6b (\citealt{Jordan2013}; \citealt{Nikolov2015}).

Although Rayleigh scattering has been seen in a number of planets to date by \emph{HST} (e.g. \citealt{Pont2013}; \citealt{Fischer2016}), it has only been seen in two hot Jupiter atmospheres from the ground (\citealt{Jordan2013}; \citealt{DiGloria2015}). Of these two ground-based detections of a Rayleigh scattering slope, one was the discovery of the slope in WASP-6b using transmission spectroscopy \citep{Jordan2013} and the other was a re-detection of the slope in HD~189733b using the chromatic Rossiter-McLaughlin effect \citep{DiGloria2015}. Whilst there have been other detections of blueward slopes from the ground, these have been considerably steeper than the expected Rayleigh scattering slopes and require further explanation (e.g. \citealt{Southworth2015}; \citealt{Parviainen2016}; \citealt{Southworth2016}). The detection in Fig. \ref{fig:trans_spec} therefore represents only the second discovery of a Rayleigh scattering slope from the ground.

\subsection{Targeted sodium search}
\label{sec:na_fits}

\begin{figure*}
\centering
\includegraphics[scale=0.5]{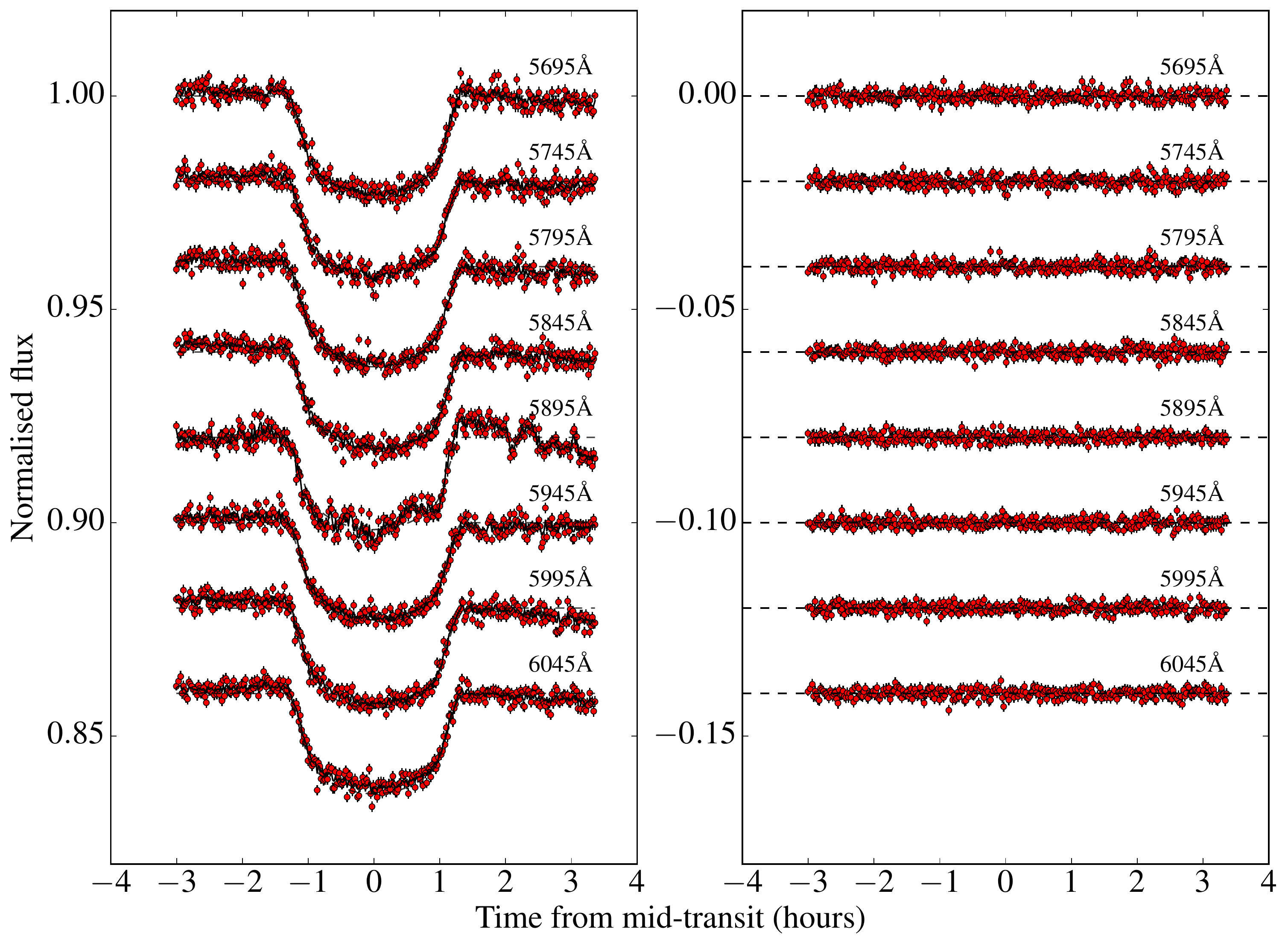}
\caption{Left-hand panel: fits to the 50\,\AA -wide wavelength bins around the sodium feature. We use a GP to model these light curves due to the significant correlated noise visible at 5895\,\AA. The dashed lines indicate the underlying \protect\cite{MandelAgol} transit model defining the mean function of the GP. Right-hand panel: the residuals to the fits in the left-hand panel after subtracting the best fitting model.}
\label{fig:na_fits}
\end{figure*}

While clouds and hazes can mask the broad wings of sodium, it is possible for the narrow line core, that arises from high altitudes, to still be visible such as in, for example, HD~189733b (\citealt{Pont2008}; \citealt{Sing2011}; \citealt{Huitson2012}; \citealt{Pont2013}) and HD~209458b (\citealt{Charbonneau2002}; \citealt{Sing2008a,Sing2008b}; \citealt{Snellen2008}; \citealt{Langland-Shula2009}; \citealt{Vidal-Madjar2011}; \citealt{Deming2013}). We performed a separate targeted search for the narrow feature of sodium using eight 50\,\AA -wide bins running from 5670 to 6070\,\AA , with one bin centred on the sodium doublet. This revealed significant red noise in the bin containing the Na feature, which can be seen in  Fig. \ref{fig:na_fits}. While the cause of this variability is unknown it could possibly be instrumental or stellar in origin. To properly account for this red noise, and to obtain robust errors for the transit depth, we again used a GP, which was implemented in the same way as when fitting the 250\,\AA-wide bins.

\begin{figure}
\centering
\includegraphics[scale=0.25]{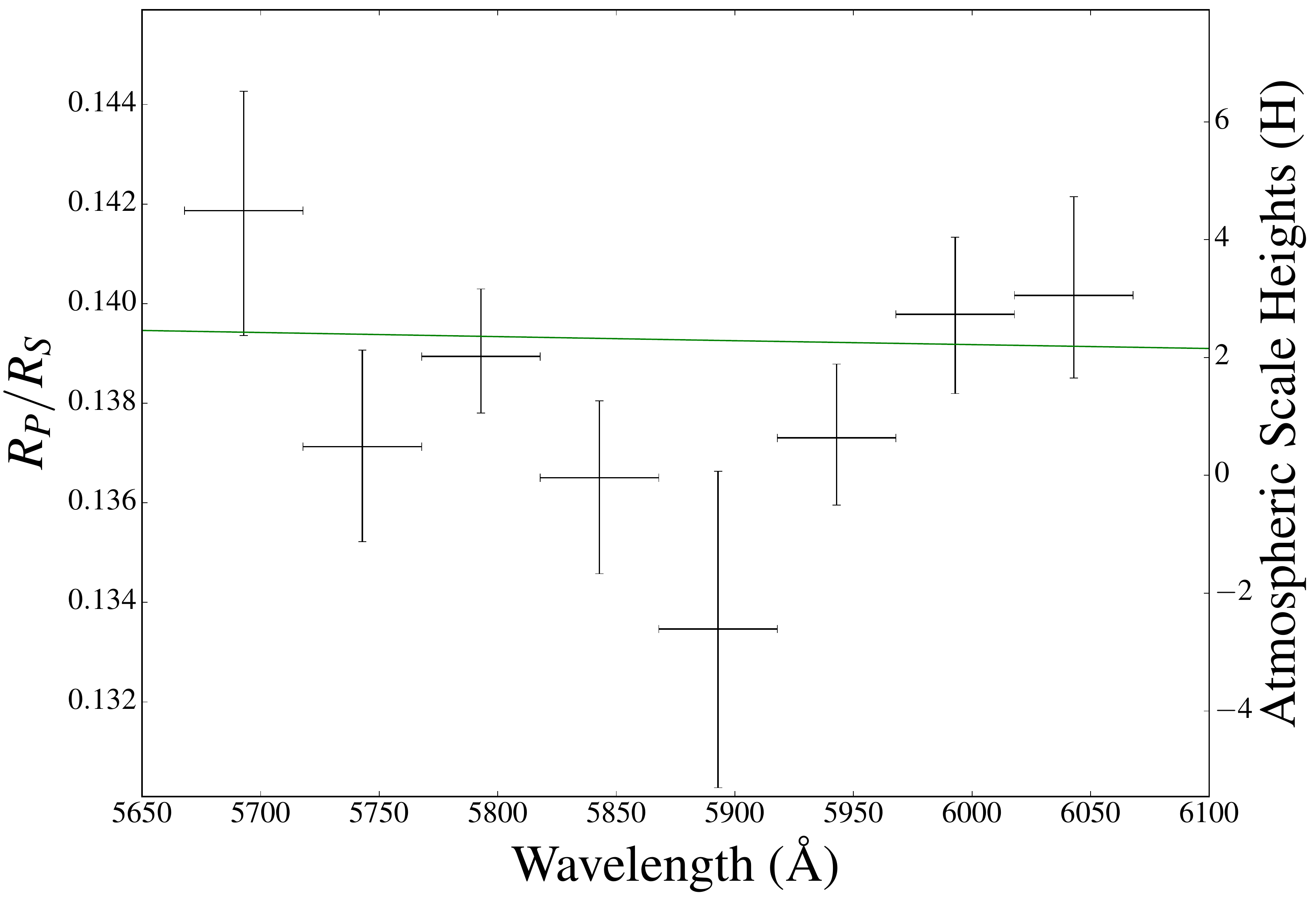}
\caption{The transmission spectrum centred around the sodium feature at a resolution of 50\,\AA, which is not detected. The green line indicates a Rayleigh slope at the equilibrium temperature of the planet.}
\label{fig:na_trans_spec}
\end{figure}

The wavelength binned fits with the GP are shown in Fig. \ref{fig:na_fits}, which displays the power of the GP to model the red noise in bin 5 whilst not overfitting the data. The transmission spectrum around the sodium feature is shown in Fig. \ref{fig:na_trans_spec}, which shows we do not detect sodium in the atmosphere at a resolution of 50\,\AA.

Whilst we attempted a targeted potassium search, unfortunately due to the potassium line's proximity to the strong telluric feature, we were unable to place constraints on the presence of this.

\subsection{Unocculted spots}

Although we do not observe any occultations of star spots within the transit of HAT-P-18b, we do need to take into account the possibility of star spots that may be present on the surface of the star but are not along the transit chord. \cite{Seeliger2015} monitored the activity of HAT-P-18 over a timespan of 12 months. They found a mean variation in the $R$-band brightness of HAT-P-18 of only $\sim0.9$\,mmag, which they find is consistent with no variation when taking their error bars into account. Therefore HAT-P-18 does not show strong activity induced photometric variation and so we do not expect unocculted activity regions to have a noticeable effect on the transmission spectrum. Nevertheless, we quantify their effects below.

Unocculted spots and plages can have the effect of inducing a slope mimicking Rayleigh scattering within the transmission spectrum and so their effect needs to be considered (\citealt{McCullough2014}; \citealt{Oshagh2014}). These unocculted spots can be accounted for by a wavelength-dependent depth correction. As with \cite{Kirk2016}, we follow the formalism of \cite{Sing2011} to make this correction.

We use \textsc{atlas9} stellar models \citep{Kurucz1993} of a star with a temperature of 4750\,K and spot temperatures with a temperature difference $\Delta$T ranging from 250 -- 1250\,K. Using this variation we assume a total dimming of 0.1 per cent at a reference wavelength of 6000\,\AA ~and use equations (4) and (5) of \cite{Sing2011} to find the correction in $R_P/R_*$ across a wavelength range spanning 4500 -- 9500\AA. We find that the effect of spots on this star is minimal as they lead to a correction in $R_P/R_*$ of between 0.00005 and 0.0001 which is a small fraction of the 1$\sigma$ error bars of our $R_P/R_*$ values. Unocculted spots cannot therefore be the cause of the blueward slope seen in the transmission spectrum (Fig. \ref{fig:trans_spec}).

\subsection{Comparison of system parameters}

With the high-quality light curves, we were able to compare the results from our white light curve fitting to those published in the literature (Table \ref{tab:system_params}), which we find are consistent with previous studies.

\begin{table*}
\caption{Comparison of system parameters to previous studies, resulting from the fits of a cubic in time polynomial and a GP to the white light curve with the limb darkening coefficients as free parameters. Although these values are consistent, the GP fit produces slightly larger errors, as expected, and so should be adopted as our conservative final values which are shown in boldface.}
\label{tab:system_params}
\centering
\begin{tabular}{l c c c c c}\hline
Parameter & This work & \protect \cite{Seeliger2015} & \protect \cite{Esposito2014} & \protect \cite{Hartman2011} \\ \hline

Cubic polynomial \\
$R_P/R_*$ & $0.1385^{+0.0010}_{-0.0011}$ & $0.1362 \pm 0.0011$ & $0.136 \pm 0.011$ &  $0.1365 \pm 0.0015$ \\
$i$ (deg.) & $88.63^{+0.12}_{-0.10}$ & $88.79 \pm 0.21$ & $88.79 \pm 0.25$ &  $88.8 \pm 0.3 $ \\
$a/R_*$ & $16.71^{+0.18}_{-0.16}$ & $17.09 \pm 0.71$  & $16.76 \pm 0.82$ &  $16.04 \pm 0.75 	$ \\
$T_c$ (BJD, days) & $2457507.59300891 \pm 0.000060 $ & $2454715.02254 \pm 0.00039$ & $2455706.7 \pm 0.7$ & $ 2454715.02174 \pm 0.00020 $ \\ 

Gaussian process \\
$R_P/R_*$ & $\boldsymbol{0.1356^{+0.0028}_{-0.0024}}$ & $0.1362 \pm 0.0011$ & $0.136 \pm 0.011$ &  $0.1365 \pm 0.0015$ \\
$i$ (deg.) & $\boldsymbol{88.53^{+0.16}_{-0.13}}$ & $88.79 \pm 0.21$ & $88.79 \pm 0.25$ &  $88.8 \pm 0.3 $ \\
$a/R_*$ & $\boldsymbol{16.39^{+0.24}_{-0.23}}$ & $17.09 \pm 0.71$  & $16.76 \pm 0.82$ &  $16.04 \pm 0.75 	$ \\
$T_c$ (BJD, days) & $\boldsymbol{2457507.59219566^{+0.000194}_{-0.000181}}$ & $2454715.02254 \pm 0.00039$ & $2455706.7 \pm 0.7$ & $ 2454715.02174 \pm 0.00020 $ \\ \hline

\end{tabular}
\end{table*}

\section{Discussion}
\label{sec:discussion}

\subsection{Transmission spectrum}

The transmission spectrum of HAT-P-18b displays a gradient rising towards the blue (Fig. \ref{fig:trans_spec}), which is consistent with Rayleigh scattering at the equilibrium temperature of the planet (852\,K). Whilst we detect the Rayleigh slope, we do not detect either the broad wings of sodium or the line core in a 50\,\AA-wide bin (Fig. \ref{fig:na_trans_spec}). This suggests that a high-altitude haze is masking the sodium feature in the atmosphere and giving rise to the Rayleigh slope, similar to, for example, WASP-12b \citep{Sing2013}, WASP-6b (\citealt{Jordan2013}; \citealt{Nikolov2014}) and HAT-P-12b \citep{Sing2016}. A haze is also preferable due to the poor fit of the clear atmosphere model in Fig. \ref{fig:trans_spec}. 

HAT-P-18b, with an equilibrium temperature of 852\,K \citep{Hartman2011}, is cooler than any of the planets studied by \cite{Sing2016} in their recent survey of 10 hot Jupiters. The nearest comparison object is HAT-P-12b \citep{Hartman2009} with an equilibrium temperature of 963\,K. For this object, \cite{Sing2016} found a haze layer leading to a strong Rayleigh scattering slope extending across the optical spectrum. They also detected potassium absorption although did not detect sodium. Whilst this might indicate some correlation of haze with temperature, \cite{Sing2016} found that the presence of clouds and hazes is not strongly dependent on temperature in their sample.

The difficulty in pinning down the relation between the presence of clouds and hazes and the planetary parameters was further highlighted in the discussion of \cite{Fischer2016} concerning the clear atmosphere of WASP-39b, which is the only hot Jupiter studied to date in which the broad wings of both sodium and potassium are visible. In this paper, the authors consider, among other parameters, the role of surface gravity and metallicity on the planet's atmosphere. If surface gravity were the dominant factor, leading to differences in the settling rates of condensates, we would expect to see condensates in the atmosphere of WASP-39b also as it has a lower surface gravity ($\log g = 2.61$, \citealt{Faedi2011}) than both HAT-P-12b ($\log g = 2.75$, \citealt{Hartman2009}) and HAT-P-18b ($\log g = 2.69$, \citealt{Hartman2011}), which is not seen in the data. If we consider metallicity, HAT-P-18 has a higher [Fe/H] of 0.1 than both HAT-P-12 (-0.29) and WASP-39 (-0.12). If we assume that the stellar metallicity is a proxy for the planet metallicity, the presence of condensates in the atmosphere of HAT-P-18b might be due to its higher [Fe/H]. The absence of condensates in the atmosphere of WASP-39b could therefore be related to its lower metallicity. This argument however breaks down when considering HAT-P-12b, which has a metal-poor host yet a condensate haze layer. It seems the presence of a haze layer is therefore not simply determined by a single parameter. 

\subsection{Limb darkening parameters}
\label{sec:limb_darkening}

\begin{figure}
\centering
\includegraphics[scale=0.25]{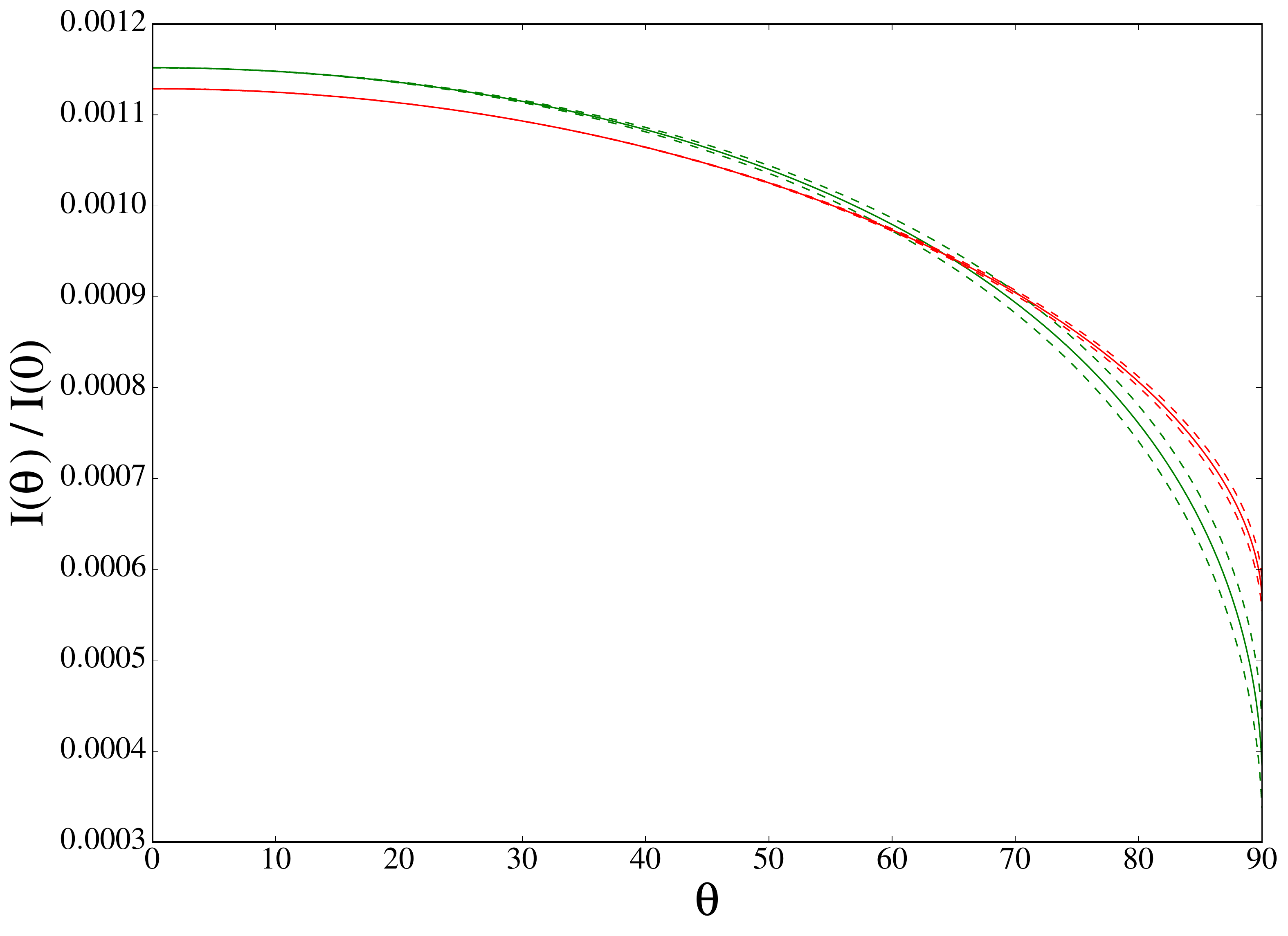}
\caption{Comparison between the stellar limb darkening profile using the predicted values for $u1$ and $u2$ (green line) with those from holding $u1$ fixed and fitting for $u2$ (red line) for wavelength bin 10. The dashed lines show the 1$\sigma$ upper and lower bounds for each. Both profiles have been normalised such that the total luminosity is equal to unity.}
\label{fig:ld_profile}
\end{figure}

When fitting our light curves with the limb darkening coefficients as free parameters, we found good agreement between the fitted and predicted values but with large error bars due to the degeneracy between $u1$ and $u2$ (Fig. \ref{fig:ldcs}, upper panel). However, fixing $u1$ at the predicted values led to $u2$ values that deviated significantly from the model (Fig. \ref{fig:ldcs}, lower panel). Fig. \ref{fig:ld_profile} shows the stellar brightness profile for one wavelength bin resulting from the use of the predicted limb darkening coefficients and the profile resulting from fixing $u1$ and fitting for $u2$. This shows that the data favours a profile with a brighter limb than the profile using the predicted limb darkening coefficients. This could indicate that we are measuring a more realistic profile of the star than the 1D model atmospheres predict. Indeed, it has been shown that limb darkening profiles resulting from 1D and 3D models do differ (e.g. \citealt{Hayek2012}; \citealt{Magic2015}). Our results highlight the value of exoplanet transits in probing stellar atmospheres in detail.

\section{Conclusions}

We have studied the transmission spectrum of the hot Jupiter HAT-P-18b using the low-resolution grism spectrograph ACAM on the WHT. We find a strong blueward scattering slope, extending across our transmission spectrum from 4750 to 9250\,\AA, consistent with Rayleigh scattering at the equilibrium temperature of the planet. We do not detect enhanced absorption around the sodium doublet, suggesting a high altitude haze is masking this feature whilst giving rise to the Rayleigh slope. We consider the effect of unocculted spots and find that the slope cannot be explained by these. This is only the second discovery of a Rayleigh scattering slope in a hot Jupiter atmosphere from the ground.

The technique of transmission spectroscopy has revealed a startling diversity in the atmospheres of hot Jupiters, some of which are clear, some cloudy and some dominated by Rayleigh scattering from a condensate haze, as we have found in HAT-P-18b. In the limited sample of systems studied to date, no clear correlations have emerged between the properties of the atmosphere and key parameters such as temperature, metallicity and surface gravity, and measurements of a wider sample of planets are desirable. Our results on HAT-P-18b demonstrate that ground-based observations are capable of detecting atmospheric opacity sources such as Rayleigh scattering. Such observations are particularly well suited to inflated hot Jupiters with large scale heights and relatively small host stars. While a relatively nearby and bright comparison star is necessary for ground-based observations, most known hot Jupiters have a suitable comparison within the length the ACAM slit (7 arcmin). For suitable targets, our observations show that 4m-class ground-based telescopes can be used to measure transmission spectra with comparable precision to \emph{HST}, and can thus play an important role in this exciting endeavour.

\section*{Acknowledgements}

We thank the anonymous referee for the careful reading of the paper and helpful suggestions. JK is supported by a Science and Technology Facilities Council (STFC) studentship. PW is supported by an STFC consolidated grant (ST/L00073). The reduced light curves presented in this work will be made available with the online journal version of this paper and at the CDS (http://cdsarc.u-strasbg.fr/). This work made use of the \textsc{astropy} \citep{astropy}, \textsc{numpy} \citep{numpy} and \textsc{matplotlib} \citep{matplotlib} \textsc{python} packages in addition to those cited within the body of the paper. The William Herschel Telescope is operated on the island of La Palma by the Isaac Newton Group in the Spanish Observatorio del Roque de los Muchachos of the Instituto de Astrof\'{i}sica de Canarias. The ACAM spectroscopy was obtained as part of W/2016A/32.




\bibliographystyle{mnras}
\bibliography{hat18_bib} 

\bsp	
\label{lastpage}
\end{document}